\documentclass[12pt]{article}
\usepackage[latin1]{inputenc}
\usepackage{amsmath, bbm, nicefrac, amsthm, booktabs, blindtext, amssymb, eurosym, amstext, setspace} %picins, english,
\usepackage[a4paper,right=3cm,left=3cm,top=2.5cm,bottom=3cm]{geometry}
\usepackage{graphicx}
\usepackage{multirow}
\usepackage[format=plain]{caption}
\usepackage[authoryear]{natbib}
\usepackage[english]{babel}
\usepackage{color, colortbl} % f�r Farben und Farben in Tabellen
\usepackage{tikz}
\usetikzlibrary{arrows,decorations.pathmorphing,backgrounds,positioning,fit}
\definecolor{dunkelgrau}{rgb}{0.8,0.8,0.8}
\definecolor{hellgrau}{rgb}{0.95,0.95,0.95}

\numberwithin{table}{section}
\numberwithin{figure}{section}

\begin{document}

\title{Optimal designs for multi-response generalized linear models with
applications in thermal spraying}

\author{  \small Holger Dette \\
\small Ruhr-Universit\"at Bochum \\
\small Fakult\"at f\"ur Mathematik \\
\small 44780 Bochum \\
\small Germany \\
{\footnotesize email: holger.dette@ruhr-uni-bochum.de}\\
\and
\small Laura Hoyden \\
\small TU Dortmund University \\
\small Department of Statistics\\
\small 44221 Dortmund \\
\small Germany \\
{\footnotesize email: hoyden@statistik.tu-dortmund.de} \\
\and
\small Sonja Kuhnt \\
\small Dortmund University \\
\small of Applied Sciences and Arts \\
\small Department of Computer Science \\
\small Postfach 105018, 44047 Dortmund \\
\small Germany \\
{\footnotesize email: sonja.kuhnt@fh-dortmund.de}\\
\and
\small  Kirsten Schorning \\
\small Ruhr-Universit\"at Bochum \\
\small Fakult\"at f\"ur Mathematik \\
\small 44780 Bochum \\
\small Germany \\
{\footnotesize email: kirsten.schorning@ruhr-uni-bochum.de}\\
}

\date{}
 \maketitle
\begin{abstract}
We consider the problem of designing experiments for
 investigating particle in-flight properties in  thermal spraying.
Observations are available on an extensive design for an initial day and thereafter in limited number for any particular day. Generalized linear models
including additional day effects are used for analyzing the process, where the models
vary with respect  to  different responses. We construct  robust $D$-optimal
designs to collect  additional data on any current day, which are efficient
for the estimation of the parameters in all models under consideration. These
designs improve a reference fractional factorial design
substantially. We also investigate designs, which
maximize the power of the test for an additional day effect.
The results are used to design additional experiments of the thermal spraying process and a comparison of the statistical analysis based on a reference design as well as on a selected D-optimal design is performed.
\end{abstract}


\maketitle
\thispagestyle{empty}
%\automark[subsection]{section}

\setcounter{tocdepth}{2}
\thispagestyle{empty}

				%%%%%%%%%%%%%%%%%%%%%%%%%%%%%%%%%%%%%%%%%%%%%%%%%%%%%%%%%%%%%%%%%%%%%%%
				%%%%%%%%%%%%%%%%%%%%%%%%%%%%%%%%%%%%%%%%%%%%%%%%%%%%%%%%%%%%%%%%%%%%%%%
				%%%%%%%%%%%%%%         Introduction       	   	  %%%%%%%%%%%%%%%%%%%%%
				%%%%%%%%%%%%%%%%%%%%%%%%%%%%%%%%%%%%%%%%%%%%%%%%%%%%%%%%%%%%%%%%%%%%%%%
				%%%%%%%%%%%%%%%%%%%%%%%%%%%%%%%%%%%%%%%%%%%%%%%%%%%%%%%%%%%%%%%%%%%%%%%

\section{Introduction}
\def\theequation{1.\arabic{equation}}
\setcounter{equation}{0}

Response surface methodology is a widely used tool to analyze the influence of experimental conditions on a response
by an adequate selection of a design and subsequent fitting of a model. It  is nowadays used in a variety of
applications, such as physics, chemistry, biology or engineering to name just a few. The precision of the estimates
can be substantially improved by the choice of an   experimental design and numerous designs which improve the statistical accuracy have been derived for the standard  linear model
[see \cite{myemon1995}, \cite{khucor1996}]. Most of the literature refers to the situation, where it is assumed that the data generating process does not change during the experiment,
however, there are many situations where this assumption may not be reasonable. We recently encountered such a situation in the context
of thermal spraying, where experiments are conducted at different days and the process is highly influenced by latent day specific effects
such as temperature or humidity. This specific application is described in detail in Section \ref{sec2}, but similar problems appear frequently
in industrial practice, whenever some latent variables change because experiments are conducted at different days. A response surface model is estimated on the basis of the available data from the first day. The experiment is continued at
another day where a limited number of additional experiments can be performed. In order to address the problem of different experimental conditions additional day effects are  included in the model. While the design of experiment for the initial day can be obtained from standard methodology,
 we are interested in an optimal design of experiment for the necessary additional experimental runs. \\
Linear
models with continuous (quite often, normally distributed)  responses as assumed in  standard response surface methodology
are  inappropriate  in the context of thermal spraying  and it is demonstrated in \cite{ThermalSprayB2012} that
generalized linear models turn out to be more suitable for describing the in-flight properties in thermal spraying. This class of  models contains the ''classical'' approach as special case and also
provides models for situations in which the response is not necessarily normal, but follows a distribution of an exponential family where the mean is modeled as a function of the predictor. Unlike the linear regression case, optimal designs then may depend on the unknown parameter value as well as the specifically chosen model components. So far, optimal designs for this situation are rarely treated in the literature and if they are  mostly with an emphasis on binary or Poisson response variables. \cite{khurietal2006} give a very nice review of the most common approaches to handle the so-called design dependency problem, namely locally optimal designs, sequential designs, Bayesian designs and quantile dispersion graphs.
 \cite{woodsetal2006} develop  a ``compromise'' design selection criterion that takes uncertainties in the parameters as well as in the link function and the predictor into account by averaging over a chosen parameter and model space. With regard to this generation of ``compromise'' designs \cite{droste2006}  present a heuristic using $K$-means clustering over local $D$-optimal designs that is robust against the mentioned uncertainties.
\\
The design problem investigated in this paper differs from the problems discussed in the literature in several perspectives. Firstly, the response in the thermal spraying process is multivariate, while the literature usually discusses designs for a univariate response. Secondly, we investigate the situation where a part of the data has been already observed on an initial day and a design is required for collecting additional data on any current day, which has good properties to estimate the parameters in the presence of a likely day-effect, describing the difference in the spraying between two days. Hence, model selection for each component of the response can be performed on the basis of the initial design, but a compromise design has to be found for the models corresponding to the different components of the response, which additionally addresses the problem of uncertainty with respect to the model parameters. While we use the $D$-optimality criterion for determining
an efficient design for estimating all parameters in the models including the additional day effect, this criterion might not be appropriate
for the purpose of detecting differences between  days. Therefore we also consider alternative criteria which are particulary designed for model discrimination.
\\
The remaining part of the paper is organized as follows. In Section \ref{sec2} we give an introduction to the problem of thermal spraying and motivate the application of generalized linear models (GLM) in this context. For the sake of transparency, we concentrate on Gamma-distributed responses
and avoid most of the general notation of GLM. Section \ref{sec4}  is devoted to optimal design problems and we discuss locally, multi-objective or compromise designs and optimal designs for identifying an additional day effect. In Section \ref{sec5} we return to the problem of designing additional experiments for the analysis of the in-flight properties in the thermal spraying problem. In particular, we demonstrate that a reference design can be substantially improved with respect to its efficiency of estimating all parameters while moderate improvements can be achieved for testing for an additional day effect.
We also develop designs with good efficiencies for both purposes.\\
The results are illustrated by designing real new  thermal spraying experiments on a different day. In particular, by performing four  additional
experiments under a reference and a  Bayesian $D$-optimal design, respectively,  it is demonstrated 
that the   Bayesian $D$-optimal design  improves the  reference design with respect to the determinant criterion for all investigated in-flight properties.
Finally, optimal designs for particular models and additional material are presented in an entire Appendix.

			%%%%%%%%%%%%%%%%%%%%%%%%%%%%%%%%%%%%%%%%%%%%%%%%%%%%%%%%%%%%%%%%%%%%%%%
			%%%%%%%%%%%%%%   Thermal Spraying	   %%%%%%%%%%%%%%%%%%
			%%%%%%%%%%%%%%%%%%%%%%%%%%%%%%%%%%%%%%%%%%%%%%%%%%%%%%%%%%%%%%%%%%%%%%%

\section{Statistical modeling of thermal spraying} \label{sec2}
\def\theequation{2.\arabic{equation}}
\setcounter{equation}{0}
Thermal spraying technology is widely used in industry to apply coatings on surfaces, aiming e.g. at better wear protection or durable medical instruments. However, due to uncontrollable factors thermal spraying processes are often lacking in reproducibility, especially if the same process is repeated on  different days. Furthermore an immediate analysis of the coating quality is usually not feasible as it requires time and results in destruction. A solution to this problem possibly lies in measuring properties of particles in flight based on the assumption that they carry the needed information of uncontrollable day effects [\cite{DVS.2010}].
We next introduce the analysed thermal spraying process and then the applied class of generalized linear models.

\subsection{Thermal spraying}

\begin{figure}[h!t]
    \centering
        \includegraphics[width=0.8\textwidth]{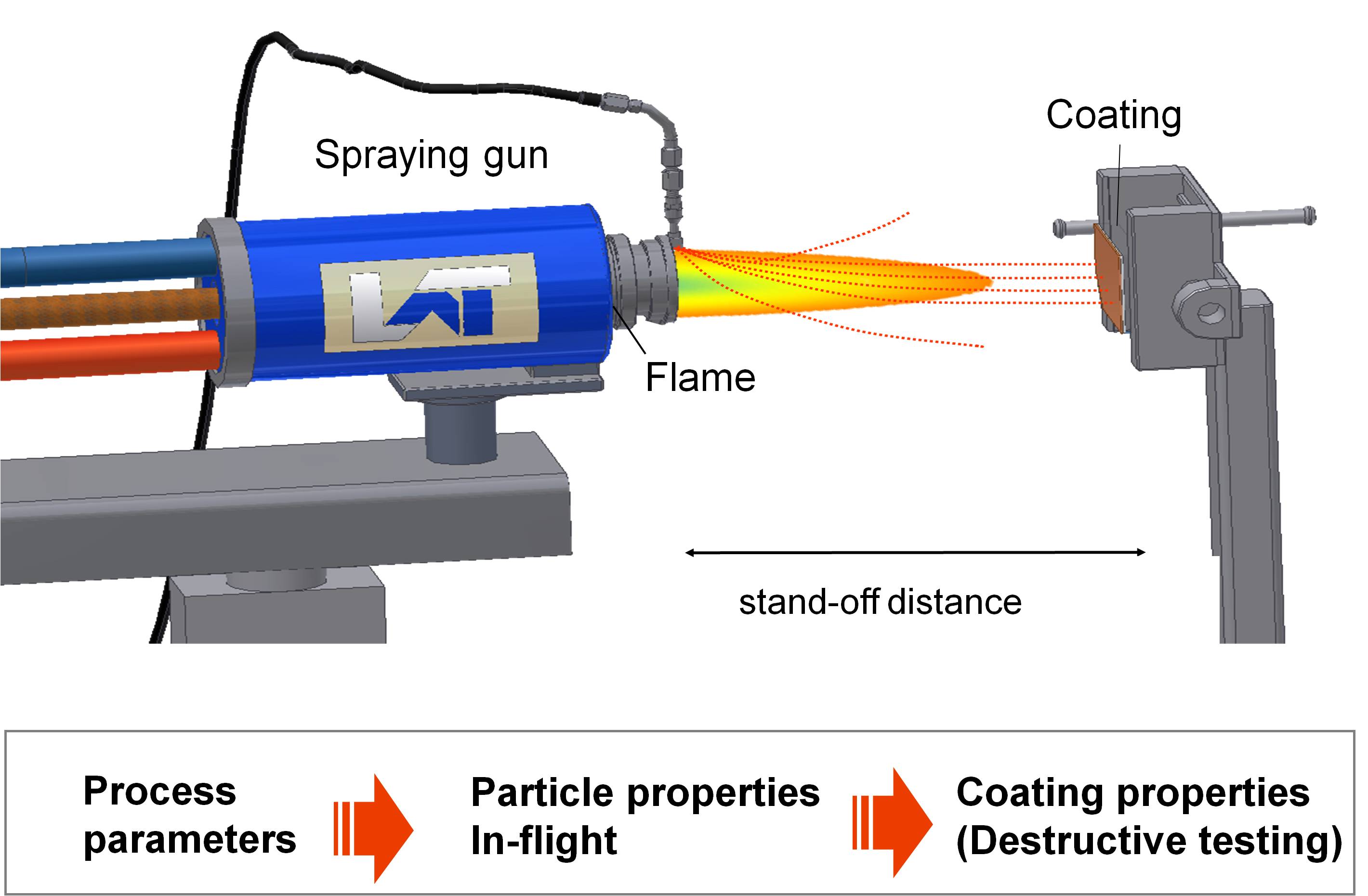}
    \caption{\emph{Thermal spraying process}}
    \label{fig:process}
\end{figure}
As application a HVOF (high-velocity oxygen-fuel spray) spraying process is regarded where WC-Co powder is melted and at high-speed applied to a surface by a spraying gun. The  influence of process parameters on  in-flight properties of the  coating powder is of interest. Figure \ref{fig:process} depicts the thermal spraying process. Preliminary  screening experiments [\cite{DVS.2010}] identify four relevant process parameters: The amount of kerosine (K) in liter per hour used, the ratio lambda of kerosine to oxygen (L) and the feeder disc velocity (FDV) as well as  the stand-off-distance (D). The last parameter describes the distance from the spraying gun to  the component which is coated and thereby also to the device measuring properties of the particles in-flight. The device measures the temperature and velocity of properties in-flight as well as flame width and flame intensity. The considered process parameters and in-flight properties are summarized in Table \ref{tab:machinesettings}.
\begin{table}[h]
\begin{center}
{\scriptsize
\begin{tabular}{l||l}
process parameters & in-flight properties \\
  \hline
stand-off-distance           ($D$)   &   temperature  \\
amount of kerosine           ($K$)   &   velocity     \\
ratio of kerosine to oxygen  ($L$)   &   flame width   \\
feeder disc velocity         ($FDV$) &   flame intensity  \\
\end{tabular}
}
\end{center}
\vspace{-.5cm}{%\small
\caption{\emph{Process parameters and in-flight properties in thermal spraying}}
\label{tab:machinesettings}}
\end{table}
Summary statistics of the in-flight measurements provide responses which have successfully been modeled by generalized linear models with Gamma distribution and different link functions based on central composite designs [\cite{ThermalSprayB2012, DAYeffectDisc2012}]. To capture the effect of unobservable day specific influences, e.g. created by  room temperature and moisture, day effects have been added to the linear predictor of the models [\cite{ThermalSprayB2012, DAYeffectDisc2012}]. These effects have to be estimated from few additional experiments on any current day. Therefore it is of high interest to determine optimal experimental designs for this specific task.

			%%%%%%%%%%%%%%%%%%%%%%%%%%%%%%%%%%%%%%%%%%%%%%%%%%%%%%%%%%%%%%%%%%%%%%%
			%%%%%%%%%%%%%%%%%%%%%%%%%%%%%%%%%%%%%%%%%%%%%%%%%%%%%%%%%%%%%%%%%%%%%%%
			%%%%%%%%%%%%%%    GLM, Fisher-Info, D-optimality	   %%%%%%%%%%%%%%%%%%
			%%%%%%%%%%%%%%%%%%%%%%%%%%%%%%%%%%%%%%%%%%%%%%%%%%%%%%%%%%%%%%%%%%%%%%%
			%%%%%%%%%%%%%%%%%%%%%%%%%%%%%%%%%%%%%%%%%%%%%%%%%%%%%%%%%%%%%%%%%%%%%%%

\subsection{Measuring information in generalized linear models}  \label{sec3}

In this section we give some background on generalized linear models which are used to model the thermal spraying process. As usually, we denote the real valued response by $Y$ and the predictor by a $q$-dimensional variable $x$. In the application $Y$ presents either the temperature, velocity, flame width or the flame intensity, while the predictor is a three- or four-dimensional variable containing some combination of the machine parameters stand-off-distance, amount of kerosine, ratio of kerosine to oxygen and feeder disc velocity. \\
Let $(Y_i,x_i)$, $i = 1,\cdots,n$, be a sample of observations where $x_i=(x_{1i},\cdots,x_{qi})^T \in\mathbb{R}^q$ are explanatory variables and $Y_i\in\mathbb{R}$ is the response at experimental condition $x_i \ (i=1,\dots,n)$. In contrast to linear models the response described by a generalized linear model may follow a distribution of the exponential family.
\cite{ThermalSprayB2012} and \cite{DAYeffectDisc2012} showed that the in-flight properties in the thermal spraying application can be adequately modeled by generalized linear models with Gamma distributed response. These models are defined by the density
\begin{equation}\label{dens}
f(y| x, \beta ) = \frac{1}{\Gamma(\nu) }   \left(\frac{\nu}{\mu}\right)^\nu  y ^{\nu-1}e^{-\frac{\nu}{\mu }y}, \quad y \geq 0,
\end{equation}
and  mean
\begin{equation}
\label{exp}
\mu = E(Y| x)=g^{-1}\left(z^T\beta\right)
 \end{equation}
where $g(\cdot )$ is an appropriate (known) link function, $z=z(x)\in\mathbb{R}^p$ is a vector
of regression functions  depending on the explanatory variables  $x$, $\beta\in\mathbb{R}^p$
denotes an unknown parameter vector and $\mu>0$ and $\nu>0$ denote the mean and shape parameter, respectively [see \cite{fahrmeir2001}].
Common link functions  for the Gamma distribution include  the identity $g(\mu)=\mu$, the canonical link
$g(\mu)=-\nicefrac{1}{\mu}$ and the log link $g(\mu)=\log(\mu)$. For the first two link functions restrictions regarding the parameter
$\beta$ have to be made such that the conditional expectation $\mu$ is non-negative. \\
If $n$ independent observations at experimental conditions $x_1,\dots,x_n$ are available and the inverse of the link function $g^{-1}$ is twice continuously differentiable, it follows by a straightforward calculation that the Fisher information matrix for the parameter $\beta$ is given by
\begin{equation} \label{infmat}
 I(\beta)= \nu^2 \sum_{i=1}^{n} w(z^T_i \beta) z_i z_i^{T} \in \mathbb{R}^{p \times p},
 \end{equation}
where the weight function is defined by
 \[ w(\mu) = ( ( \log g^{-1} (\mu))')^2= \frac {1}{(g'(g^{-1}(\mu)) g^{-1}(\mu))^2}.
 \]  The covariance matrix of the maximum likelihood estimator for the parameter $\beta$ can be approximated by   the inverse of the information matrix $I(\beta)$. Note that for the different link functions the corresponding information matrices differ only with respect to the weight $w(\mu)$, and the weights corresponding to the  Gamma distribution for the named link functions are shown in Table \ref{tab:fisherinfo_gamma}.\\
\vspace{-0.5cm}
\begin{table}[!h]
\centering
{\normalsize
\begin{tabular}{l|c}
link function $g(\cdot)$   			& weight in \eqref{infmat}   \\ \hline
$g(\mu)=\mu$								&  ${1}/{(z_i^{T}\beta)^2}$ \\
$g(\mu)= \nicefrac{1}{\mu}$ & $ {1}/{(z_i^{T}\beta)^2}$ \\
$g(\mu)=\log(\mu)$					& $1$ \\
\end{tabular}}
\vspace{-.2cm}
\caption{\it{Weights in the information matrix \eqref{infmat} for the Gamma distribution with identity, canonical and log link}}
\label{tab:fisherinfo_gamma}
\end{table}
\vspace{-0.5cm}

In each case the information matrix depends on the sample size $n$, the link function $g$, the vector of regression functions  $z(x)$ and especially on the  parameter $\beta$.
Throughout this paper we consider a quadratic response function for $g(E[Y|x])$, that is
\begin{equation} \label{resp}
z^T\beta= \beta_0 + \sum^q_{i=1} \beta_i x_i + \sum^q_{i=1} \sum^q_{j\geq i} \beta_{ij} x_i x_j.
\end{equation}

			%%%%%%%%%%%%%%%%%%%%%%%%%%%%%%%%%%%%%%%%%%%%%%%%%%%%%%%%%%%%%%%%%%%%%%%
			%%%%%%%%%%%%%%%%%%%%%%%%%%%%%%%%%%%%%%%%%%%%%%%%%%%%%%%%%%%%%%%%%%%%%%%
			%%%%%%%%%%%%%%    GLM, Fisher-Info, D-optimality	   %%%%%%%%%%%%%%%%%%
			%%%%%%%%%%%%%%%%%%%%%%%%%%%%%%%%%%%%%%%%%%%%%%%%%%%%%%%%%%%%%%%%%%%%%%%
			%%%%%%%%%%%%%%%%%%%%%%%%%%%%%%%%%%%%%%%%%%%%%%%%%%%%%%%%%%%%%%%%%%%%%%%

\section{Optimal designs for generalized linear models}  \label{sec4}
\def\theequation{3.\arabic{equation}}
\setcounter{equation}{0}

Optimal designs maximize a functional, say $\Phi$, of the Fisher information matrix with respect to the choice of the experimental conditions $x_1,\dots,x_n$, and numerous criteria have been proposed in the literature to
discriminate between competing designs [see \cite{pukelsheim2006}].
The  commonly used optimality criteria (such as the $D$-, $A$- or $E$-optimality criterion) are positively
homogenous, that is $\Phi(\lambda I(\beta))=\lambda \Phi (I(\beta))$ whenever $\lambda \geq 0$ [see \cite{pukelsheim2006}]. Consequently, an optimal design maximizing a functional of the Fisher information matrix \eqref{infmat} will not
depend on the parameter $\nu$, but it will depend on the parameter $\beta$. Designs depending on unknown parameters of the model
 are called locally optimal designs and were at first discussed by
 \cite{chernoff1953}. Since this fundamental paper many authors have worked in the construction of locally
 optimal designs. We refer to some recent work in this direction by
\cite{yangstuf2009}, \cite{yang2010} and \cite{detmel2011}, who discuss admissible classes of locally optimal designs for nonlinear regression models with a one-dimensional predictor.\\
In situations where  preliminary knowledge regarding the unknown parameters of a nonlinear model is available,
the application of
locally optimal designs is well justified. A typical example are phase II clinical dose finding trials, where some useful knowledge is
already available from phase I [see \cite{debrpepi2008}]. A further example is given by the thermal spraying problem introduced in Section \ref{sec2}. Here a couple of experiments were already performed on the basis of a central composite design, and new experiments
have to be planned for further investigations. On the basis of the available observations parameter estimates and standard deviations are
available, which can be used in the corresponding local  optimality criteria. Locally $D$-optimal designs for the generalized linear model introduced by \eqref{dens}, \eqref{exp}, \eqref{resp} will be defined in Section \ref{sec41} and discussed in \ref{sec51}.
\\
 On the other hand,  locally optimal designs are often used  as  benchmarks for  commonly proposed designs (see also the discussion in Section \ref{sec5}). Moreover, they are the basis for
  more sophisticated design strategies, which require less
precise knowledge about the model parameters,
 such as sequential, Bayesian or standardized maximin optimality criteria  [see  \cite{pronwalt1985},  \cite{chaver1995} and
\cite{dette1997}   among others]. Optimal designs with respect to the latter  criteria are  called robust designs and  will be discussed in Section
\ref{sec42}. Finally, optimal designs for investigating the existence of an additional day effect are introduced in Section \ref{sec43}.

\subsection{Locally $D$-Optimal designs} \label{sec41}
As \cite{myemovinn2002}  point out, the $D$-optimality criterion is a commonly used design selection criterion especially for industrial experiments.
To be precise, consider a   link function $g$ and   a  regression model of the form \eqref{resp} defined with corresponding vector  $z=z(x)$ and
 parameter $\beta$. We collect the model information in the vector $s=(g,z,\beta)$.
In order to reflect the dependency of the Fisher information  matrix   in \eqref{infmat}
on the design and on the particular model specified by the link function $g$ and corresponding parameter $\beta$  we introduce the notation
\begin{equation} \label{fish1}
I(\mathbf{X}, s ) = \sum_{i=1}^{n}w (z_i,\beta)  z_iz_i^{'}
\end{equation}
for the Fisher information matrix,
where $\mathbf{X}=(x_1,\cdots,x_n)$   denotes the design  and $z_i =z(x_i)$ ($i=1,\ldots , n$).  Following \cite{chernoff1953}  we call a design
$\mathbf{X}^*_s$  locally $D$-optimal
if it maximizes   the determinant of the Fisher information matrix
 \begin{eqnarray}
\Phi_D (\mathbf{X} |s  ) = \left|I(\mathbf{X},s)\right|^{1/p(s)},
\label{dopt}
\end{eqnarray}
where $p(s)$ denotes the number of parameters in model $s$. Note that the locally $D$-optimal design depends on the link function $g$, the model $z$ and the corresponding unknown parameter vector $\beta$, which justifies our notation $\mathbf{X}^*_s$ ($s=(g, z, \beta)$).
It is usually assumed that information regarding the unknown parameter in a specific fixed model is available
[see for example  \cite{fortorwu1992},  \cite{biedetzhu2006},  \cite{fangheda2008},  \cite{detkisbevbre2010} among many others]. Locally $D$-optimal designs  (and other optimal designs with respect to local optimality criteria)  have been criticized because of their dependence on the specific choice of  the parameter $\beta$. However, there are numerous situations where  preliminary knowledge
regarding the unknown parameters  is available, such that  the application of
locally optimal designs is well justified (see the discussion at the beginning of this section).
  A further common criticism of the criterion \eqref{dopt} is that it requires the specification of the model and the link function and there are several
  situations where a design for a specific model is not efficient for an alternative competing  model [see \cite{debrpepi2008}]. For example in the case of thermal spraying, different models turn out to be appropriate for analyzing the temperature, velocity, flame width and intensity but data can only be collected according to one design. In Section \ref{sec51} we demonstrate that a locally D-optimal design for analyzing one particular in-flight property might be inefficient for analyzing another one.\\
  In the following sections we briefly discuss  different approaches to find   $D$-optimal designs which are less sensitive with respect to a misspecification of link, model and parameter vector $\beta$.

     %%%%%%%%%%%%%%%%%%%%%%%%%%%%%%%%%%%%%%%%%%%%%%%%%%%%%%
     %%%%%%%%%%%%%%%%%%%%%%%%%%%%%%%%%%%%%%%%%%%%%%%%%%%%%%
     %%%%%%%%%%%%%%%%%%%%%%%%%%%%%%%%%%%%%%%%%%%%%%%%%%%%%%

\subsection{Multi-objective designs} \label{sec42}
The problem of addressing model uncertainty (with respect to the form of the regression function or prior information regarding the unknown parameter) has a long history. \cite{laeuter1974a}   proposed a criterion which is based on a product of the determinants of the information matrices in the various models under consideration and yields designs which are efficient for a class of given models. \cite{laustu1985}  and \cite{dette1990} determine optimal designs with respect to L\"{a}uter's criterion for a class of trigonometric and polynomial regression models, respectively.
In the case  where the form of the model is fixed and  there is uncertainty about the nonlinear parameter
\cite{laeuter1974b} and \cite{challarn1989}   propose a Bayesian $D$-optimality criterion which maximizes an expected value of the $D$-optimality criterion with respect to a prior distribution for the unknown parameter [see also \cite{pronwalt1985}, who call  the corresponding designs robust designs, or \cite{chaver1995}  for comprehensive reviews of this approach]. Since its introduction Bayesian optimal designs have found considerable attention in the literature [see \cite{haines1995}, \cite{mukhai1995}, \cite{detneu1997}, \cite{hancha2004}   among others].
\cite{biedetpep2006}  determined efficient designs for binary response models, when there is uncertainty about the form of the link function (e.g.\ Probit or Logit model) and the parameters. Recently,
\cite{woodsetal2006} used this approach for finding   $D$-optimal designs in the case of uncertainty   concerning the parameter vector $\beta$ as well as the linear predictor $\eta=z^{'}\beta$ and the link function $g(\cdot)$. For this purpose these authors propose a multi-objective criterion [see \cite{cookwong1994}] for the selection of a design. Most of the optimality criteria in these references are based on the  average of given optimality criteria $\Phi(\mathbf{X}|s)$ (such as the $D$-optimality criterion) over the space $\mathcal{M}$ of the possible models, which takes the model uncertainty into account. In the present context the elements of the set $\mathcal{M}$ are of the form $s=(g,z,\beta)$ corresponding to uncertainty with respect to the link function $g$, the regression function $z=z(x)$ and the parameter $\beta$.\\
To be precise, let $\mathcal{G}$ denote a class of possible link functions. For each $g \in \mathcal{G}$ let $\mathcal{N}_g$ denote a class of vector-valued functions $z(x)$ and finally define for each pair $(g,z)$ with $z \in \mathcal{N}_g$ a parameter space $\mathcal{B}_{g,z}$.
 %(, that are depending on $\beta$, $\eta$ and $g$)
With
$\mathcal{M}=\{(g,z,\beta):g\in\mathcal{G},z \in \mathcal{N}_g ,\beta\in  \mathcal{B}_{g,z} \}$ %(where $\mathcal{G}$, $(\mathcal{N}|g)$ and $(\mathcal{B}|g,\eta)$  are the sets of possible link functions, linear predictors and parameter vectors,)
the robust optimality criterion is given by
\begin{equation} \label{critbay}
\Phi_{B} (\mathbf{X}|\mathcal{M})= \int_{ \mathcal{M}} \mbox{eff}_D   ( \mathbf{X}| s) dh_1(\beta|g,z)dh_2(z|g)dh_3(g),
\end{equation}
where  the efficiency is defined by
\begin{equation} \label{eff}
\text{eff} _D(\mathbf{X}| s  )=\frac{\Phi_D (\mathbf{X}| s )}{\Phi_D (\mathbf{X}^*_s| s  )},
\end{equation}
$\mathbf{X}^*_s$ is the locally $D$-optimal design for model $s\in \mathcal{M}$ and $h_1$, $h_2$ and $h_3$ represent cumulative distribution functions reflecting the importance of the particular constellation $(g,z,\beta)$. \\
As an alternative to the Bayesian criterion
\cite{dette1997}  proposed a standardized maximin $D$-optimality criterion, which determines a design maximizing the worst efficiency over the class $\mathcal{M}$ of possible models [see also \cite{muepaz1998b}]. Since its introduction this criterion has found considerable attention in the literature. To be precise, assume that $\mathcal{M} $ is a set of possible values $s=(g,z,\beta)$
   for the link function, model and  parameter vector and recall the definition of  the relative efficiency of the design $\mathbf{X}$ with respect to the
   locally optimal  design $\mathbf{X}^*_s$ defined by
\eqref{eff}. The standardized maximin optimal design $\mathbf{X}^*$ is defined as the solution of the  optimization problem
\[\max_{\mathbf{X}}\min_{s  \in \mathcal{M}} \text{eff}_D\left(\mathbf{X}| s \right).\]
Therefore this design   maximizes the minimal relative efficiency calculated over the set $\mathcal{M}$, and it can be expected that such a design has reasonable efficiency for any choice of the parameter $s \in \mathcal{M}$.
%Chipman and Welch (1996) assume the approach to be robust over the set of plausible parameter values. But also suggest to extend the set of designs ${\mathbf{X_1},\cdots,\mathbf{X_k}}$ to receive a design which provides a good overall efficiency although it is not D-optimal for any of the $\beta_k\in B$.
\\
Standardized maximin optimal designs are extremely difficult to find and for this reason we will mainly consider optimal designs with respect to the Bayesian-type criterion \eqref{critbay}. Some explicit results for models with a one-dimensional predictor can be found in \cite{imhof2001} as well as  \cite{dethaiimh2007}.

     %%%%%%%%%%%%%%%%%%%%%%%%%%%%%%%%%%%%%%%%%%%%%%%%%%%%%%
     %%%%%%%%%%%%%%%%%%%%%%%%%%%%%%%%%%%%%%%%%%%%%%%%%%%%%%
     %%%%%%%%%%%%%%%%%%%%%%%%%%%%%%%%%%%%%%%%%%%%%%%%%%%%%%

			%%%%%%%%%%%%%%%%%%%%%%%%%%%%%%%%%%%%%%%%%%%%%%%%%%%%%%%%%%%%%%%%%%%%%%%
			%%%%%%%%%%%%%%%%%%%%%%%%%%%%%%%%%%%%%%%%%%%%%%%%%%%%%%%%%%%%%%%%%%%%%%%
			%%%%%%%%    Day-effect model under gamma distribution	   %%%%%%%%%%%%%%
			%%%%%%%%%%%%%%%%%%%%%%%%%%%%%%%%%%%%%%%%%%%%%%%%%%%%%%%%%%%%%%%%%%%%%%%
			%%%%%%%%%%%%%%%%%%%%%%%%%%%%%%%%%%%%%%%%%%%%%%%%%%%%%%%%%%%%%%%%%%%%%%%

\subsection{Design criteria in the presence of an additional day-effect} \label{sec43}

Recall the motivating example discussed at the end of Section \ref{sec2}, where observations are taken at two
different days. In order to address this situation in the generalized linear model we replace
the regression model $z(x)$ and the parameter $\beta$ in \eqref{exp} by  the vectors
 $$
 z^*(x,t) = (z(x)^T,t)^T \ ;  \quad
 \beta^* = (\beta^T,\gamma)^T
$$
 respectively, where the parameter $t$ can attain the values $0$ and $1$ corresponding to different
 experimental conditions caused by a possible  day effect. Thus the expected response at a particular experimental condition satisfies
 \begin{eqnarray}\label{daymod}
 g(E[Y|x])= \left \{
 \begin{array}{lll}
 z^T(x)\beta & \mbox{if} & t=0 \\
 \gamma + z^T(x)\beta & \mbox{if} & t=1.
 \end{array} \right.
 \end{eqnarray}
 We assume that $n$ observations
are taken at the initial day at experimental conditions $x_1,\ldots , x_n$.
This corresponds to the choice $t=0$ and a generalized linear model without the day
effect $\gamma$ is fitted to the data.  Additional experiments can be made at any further day at experimental
conditions $x_{n+1},\ldots , x_{n+m}$
which corresponds to
the choice $t=1$. The Fisher information for a specific
model, weight function  (corresponding to the generalized linear model) and  parameter is then
given by
\begin{equation} \label{fish2}
I(\mathbf{X}, s ) = \sum_{i=1}^{n+m}w ({z_i^*}^T  \beta^*)  z_i^*{z^*_i}^{T}   \in \mathbb{R}^{p(s)+1 \times p(s)+1}
\end{equation}
where  $z_i^*= z(x_i,t_i) $ denotes the vector of regression functions corresponding to the
$i$-th observation ($i=1, \ldots, n+m$) and the weight function is defined by
$$
\frac{1}{\left({z^*_i}^{T}\beta^*\right)^2}~,~\frac{1}{\left({z^*_i}^{T}\beta^*\right)^2}~,~~
1
$$
for the identity, inverse and log-link, respectively.
Note that in the matrix
$${\bf{X}}=({\bf{X}}^{(1)}, {\bf{X}}^{(2)})=(x_1,\dots,x_n,x_{n+1},\dots,x_{n+m})$$
the elements   in the matrix ${\bf{X}}^{(1)}=(x_1,\dots,x_n)$ are fixed (because they correspond to observations from the initial day) and the criteria are optimized with respect to the experimental conditions ${\bf{X}}^{(2)}=(x_{n+1},\dots,x_{n+m})$ for the experiments at a different day. We reflect this fact by the notation
\begin{eqnarray} \label{crit2}
\Phi_D({\bf X}^{(2)}|s)= \Phi_D(({\bf X}^{(1)},{\bf X}^{(2)})|s) \label{Dday} \\
\Phi_B({\bf X}^{(2)}|\mathcal{M})= \Phi_B(({\bf X}^{(1)},{\bf X}^{(2)})|\mathcal{M}) \label{Bday}
\end{eqnarray}
for the criteria \eqref{dopt} and \eqref{critbay}, respectively. The corresponding locally optimal designs are denoted by ${\mathbf{X}_s^*}^{(2)}$ and the analogue of the efficiency \eqref{eff} is given by
\begin{equation} \label{effday}
\mbox{eff}_D({\bf {X}}^{(2)}|s) =\left(\frac{|I(({\bf X}^{(1)},{\bf X}^{(2)}),s)|}{|I(({\bf X}^{(1)},{\bf X}_s^{*(2)}),s)|}\right)^{1/p(s)+1},
\end{equation}
where the Fisher information matrix $I$ is defined in \eqref{fish2}. The $D$-optimality criterion is well justified if the main goal is to estimate all parameters in the presence of such day effects.\\
On the other hand other optimality criteria should be used if the only goal of the experiment is the investigation of an additional day effect. For this purpose a likelihood ratio test for the  hypothesis
\begin{equation} \label{h0}
H_0: \gamma=0
\end{equation}
on the basis of all $n+m$ observations is usually performed. Standard results on the asymptotic properties of the
likelihood ratio test show that the power of the test for the hypothesis \eqref{h0}  in a model $s=(g,z,\beta)$
is an increasing function of the quantity
 \begin{eqnarray}
\Phi_{D_1} (\mathbf{X}^{(2)}|s ) =  (e_{p(s)+1}^T  I^{-1}(\mathbf{X},s)   e_{p(s)+1})^{-1}
\label{dopt1}
\end{eqnarray}
where $\mathbf{X}=(\mathbf{X}^{(1)}, \mathbf{X}^{(2)}), \ \mathbf{X}^{(1)}=(x_1,\dots,x_n), \ \mathbf{X}^{(2)}=(x_{n+1},\dots,x_{n+m})$ and
 $ e_{p(s)+1} = (0,\ldots ,0,1)^T $ denotes the $(p(s)+1)$-th unit vector in $ \mathbb{R}^{p(s)+1}$ [see \cite{debrpepi2008}]. Consequently, an optimal design
for investigating the existence of an additional day effect if a particular model $s=(g,z,\beta)$ is used for the data analysis maximizes
the function $\Phi_{D_1} (\mathbf{X}^{(2)}|s ) $ with respect to the choice of the experimental conditions $\mathbf{X}^{(2)}=(x_{n+1},\ldots , x_{n+m})$ for the
 $m$ observations taken at any further day. The criterion defined by  (\ref{dopt1}) is called $D_1$-optimality criterion in the literature. $D_1$-optimal designs have been studied by several authors in the context of linear and nonlinear regression models [see \cite{studden1980}, \cite{detmelwong2005} or \cite{detkisbevbre2010} among others], but less work can be found on $D_1$-optimal designs for generalized linear models.
\\
 In order to address uncertainty with respect to the model assumptions we denote by $\mathbf{X}^{*(2)}_s$ the locally $D_1$-optimal design maximizing the criterion defined in \eqref{dopt1} and define the $D_1$-efficiency  of a design $\mathbf{X}=(\mathbf{X}^{(1)},\mathbf{X}^{(2)})$
 in model $s=(g,z,\beta)$ by
 \begin{equation} \label{eff1}
\text{eff}_{D_1}(\mathbf{X}^{(2)}| s  )=\frac{\Phi_{D_1} (\mathbf{X}^{(2)}| s )}{\Phi_{D_1} (\mathbf{X}^{*(2)}_s| s  )}.
\end{equation}
 The Bayesian $D_1$-optimality criterion is finally defined by
 \begin{equation} \label{critbayD1}
\Phi_{B_1} (\mathbf{X}^{(2)}|\mathcal{M})= \int_{ \mathcal{M}} \mbox{eff}_{D_1}   ( \mathbf{X}^{(2)}| s) dh_1(\beta|g,z)dh_2(z|g)dh_3(g),
\end{equation}
where    $h_1$, $h_2$ and $h_3$ represent again cumulative distribution functions reflecting the importance of
the particular constellation $(g,z,\beta)$.  Criteria of this type have been discussed by several authors
in the case of linear regression models    [see \cite{dette1994a}, \cite{dethal1998}].

\section{Optimal designs for thermal spraying} \label{sec5}
\def\theequation{4.\arabic{equation}}
\setcounter{equation}{0}

We return to the problem of designing additional experiments for the  thermal spraying process. In the application the design space for each variable is the interval $[-2,2]$ and $30$ observations have already been made on the basis of a central composite design $\mathbf{X}_C^{(1)}$
(see Table  \ref{tab:CCD}  in Appendix \ref{secbestfits}) while a small number of additional experiments, e.g. four, are to be conducted for the investigation of an additional day effect.
For each response (temperature, velocity, flame width, flame intensity) the data from the initial day has been used to identify a  generalized linear model  in the class of all models with the three link functions
specified in Section \ref{sec3}  and different forms for the vector $z$ on the basis of the  BIC. The corresponding results are
listed in Table \ref{tab:Properties}.
For each  response the parameter estimates corresponding to the model chosen by the BIC
are shown in Table \ref{tab:bmtemai} in Appendix \ref{secbestfits}. 
\begin{table}[t!]
\centering
{\scriptsize
\begin{tabular}{rccccc}
\hline
 & temperature %&  temperature (FDV)
  & velocity & flame width & flame intensity \\
\hline
 Main effects & $L, K, D$ %& $L, K, D, FDV$
 &$L, K, D, FDV$ & $L, K, D, FDV$ & $L, K, D, FDV$ \\
 Squared effects & $K^2$ %& $K^2$
 & $K^2$ & $K^2$ & $L^2, K^2, FDV^2$ \\
 Interaction terms & -- %& --
 & $L \cdot K$ & -- & $D \cdot FDV$ \\
 \hline
Link & identity %& identity
& logistic & inverse & identity \\
\hline
\hline
 BIC & 245.744 %&  245.745
 & 196.979 & 99.749 & 106.148 \\
 %Link function & Gamma (identity) & Gamma (logistic) & Gamma (inverse) & Gamma (identity) \\
% Link & logistic & identity & inverse & identity \\
\hline
\end{tabular}}
\vspace{-.2cm}
\caption{\it The generalized linear models chosen by the BIC for the four responses observed in the thermal
spraying process.}
%\caption{Properties of the chosen GLMs with Gamma-distributed response variable}
\label{tab:Properties}
\end{table}
For example,
for the temperature the BIC selects the generalized linear model with gamma distribution
and identity link where the linear part of the model  is given by
%$$ (mit FDV)
%z^T(x)\beta= \beta_0 + \beta_1 L +\beta_2K + \beta_3D + \beta_4 FDV + \beta_5 K^2  .
%$$
$$ %(ohne FDV)
z^T(x)\beta= \beta_0 + \beta_1 L +\beta_2K + \beta_3D + \beta_4 K^2  .
$$
The estimated values of the parameters $(\beta_0,\dots,\beta_4)$  can be obtained  from Table \ref{tab:bmtemai}. For the investigation of the existence of an
additive  day effect a reference design $\mathbf{X}_R^{(2)} = (x_{31},\dots,x_{34})$ for the four additional experiments was proposed, which is shown in
 Table \ref{locd1ref}. In order to investigate the efficiency of this design we have calculated the
 best locally $D$-optimal designs for the  models which were identified by the BIC for modeling
 the four responses with an additional day effect. These designs require the specification of the unknown parameters and
 we used the available information from the first $30$ experiments of the first day to estimate  $\beta$ (see Table \ref{tab:bmtemai}),
 while the parameter $\gamma$ for the additional day effect was chosen (on the basis of information from similar experiments) as $\gamma =-16$, $\gamma =0.01$, $\gamma= 0.002$ and $\gamma= 0.09$
in the models  for  temperature, velocity, flame width and flame intensity, respectively.\\
All designs presented in this section are calculated by Particle Swarm Optimization (PSO) which was introduced by \cite{eberken1995}. We also refer to the monographs \cite{clerc2006} and \cite{yangXS2010} for the general methodology.

\subsection{$D$-optimal designs} \label{sec51}
 The locally $D$-optimal designs
  are shown in Table \ref{tab:locoptD} in  Appendix \ref{secdesigns2}, while the corresponding $D$-efficiencies
  $$
  \mbox{eff}_D (\mathbf{X}^{(2)}_R|s) =  \left( {| I(\mathbf{X}_R,s) |  \over |I({\mathbf{X}^*_s},s)|} \right)^{1/(p(s)+1)}
  $$
   for the designs $\mathbf{X}_R=(\mathbf{X}_C^{(1)}, \mathbf{X}_R^{(2)})$ and $\mathbf{X}_s^*=(\mathbf{X}_C^{(1)}, \mathbf{X}_s^{*(2)})$ are depicted in the first row of Table \ref{efficlocDopt}. Here $I({\bf{X}},s)$ is the Fisher information in the generalized linear model including the day effects and  $p(s)+1$ denotes the number of parameters in the corresponding model where $p(s)$ parameters appear in regression function $z^T(x)\beta$. We observe that
  for each  type of response the corresponding locally $D$-optimal design yields a substantial improvement of the reference design.
  The efficiency of the reference design varies between $65\%$ - $80\%$.
 \begin{table*}
	\centering
	{\scriptsize
		\begin{tabular}{|c|c|c|c|}
		\hline
			  temperature & velocity & flame width & flame intensity \\
			 \hline
			 \hline
			 $80.03 \% $ & 71.13\% & 68.11\% & 64.85\% \\
			  81.62\% & 72.94\% & 74.48\% & 69.91\% \\
			  79.56\% & 74.58\% & 74.58\% & 72.21\% \\
			\hline
%			Link	& identity & logistic & inverse & identity\\
%			\hline
		\end{tabular}}
		\vspace{-.1cm}
	\caption{\it First row: $D$-efficiencies of the reference design. Second row: $D$-efficiencies of the reference design $\mathbf{X}^{(2)}_R$
	with respect to
	the design $\mathbf{X}_B^{*(2)}$ maximizing the multi objective criterion \eqref{Bday}, where $\gamma$ has been fixed.
	Third row: $D$-efficiencies of the reference design $\mathbf{X}^{(2)}_R$
	with respect to
	the design $\mathbf{X}_B^{*(2)}$ maximizing the multi objective $D$-criterion \eqref{Bday}, where uncertainty with respect to the parameter $\gamma $ has been addressed.}
	\label{efficlocDopt}
	\end{table*}
	
Because an important goal of the experiment is to answer the question of additional day effects we display in Table \ref{Ddayeff} the $D_1$-efficiencies
\begin{equation} \label{effsR}
\mbox{eff}_{D_1} (\mathbf{X}^{(2)}_R, \mathbf{X}_s^{*(2)}|s)= {\Phi_{D_1}(\mathbf{X}_R^{(2)}|s) \over \Phi_{D_1}(\mathbf{X}_s^{*(2)}|s)}
\end{equation}
of the reference design $\mathbf{X}_R=(\mathbf{X}^{(1)}_C, \mathbf{X}^{(2)}_R)$ with respect to the locally $D$-optimal design $(\mathbf{X}^{(1)}_C, \mathbf{X}^{(*2)}_s)$ for estimating the parameter $\gamma$. Most of the $D_1$-efficiencies of the locally $D$-optimal designs are larger than $100 \%$ compared to the reference design. The locally $D$-optimal design only performs better if flame width is concerned. Summarizing this means that the locally $D$-optimal design does not yield an improvement of the reference design when the only goal of the experiment is a most precise estimation of the additional day effect. Therefore we also investigate locally $D_1$-optimal designs in Section \ref{sec52} in order to optimize the power of the test for an additional day effect.

 \begin{table*}[t!]
	\centering
	{\scriptsize
		\begin{tabular}{|c|c|c|c|}
		\hline
			  temperature & velocity & flame width & flame intensity \\
			 \hline
			 \hline
			  100.17\% & 195.57\% & 76.31\% & 202.41\% \\
			   132.76\% & 129.17\% & 85.39\% & 515.47\% \\
			  132.76\% & 131.51\% &  87.92\% & 508.25\% \\
			  			\hline
%			Link	& identity & logistic & inverse & identity\\
%			\hline
		\end{tabular}}
		\vspace{-.1cm}
		\caption{\it First row: $D_1$-efficiencies of the reference design with respect to the locally D-optimal designs for estimating the parameter $\gamma$ (see formula \eqref{effsR}). Second row: $D_1$-efficiencies of the reference design $\mathbf{X}^{(2)}_R$
	with respect
	to the design $\mathbf{X}_B^{*(2)}$ maximizing the multi objective $D$-criterion \eqref{Bday}, where $\gamma$ has been fixed.
	Third row: $D_1$-efficiencies of the reference design $\mathbf{X}^{(2)}_R$
	with respect
	to the design $\mathbf{X}_B^{*(2)}$ maximizing the multi objective $D$-criterion \eqref{Bday}, where uncertainty with respect to the parameter $\gamma $ has been addressed.}
	\label{Ddayeff}
\end{table*}

\begin{table*}[t!]
	\centering
	{\scriptsize
		\begin{tabular}{|c||c|c|c|c|}
		\hline
		Locally $D$-optimal & \multicolumn{4}{c|}{Model}\\
		 \cline{2-5}
			 design for & temperature & velocity & flame width & flame intensity \\
			 \hline
			 \hline
			temperature &  100.00\%& 85.19\% &80.88\% & 72.91\% \\
			   velocity &96.15\% & 100.00\% & 84.81\% & 93.05\% \\
			  flame width & 91.79\% & 84.29\% & 100.00\% & 73.05\%\\
			flame intensity & 96.19\%  & 87.93\% & 65.77\% & 100.00\% \\
			\hline
%			Link	& identity & logistic & inverse & identity\\
%			\hline
		\end{tabular}}
		\vspace{-.2cm}
		\caption{\it The efficiencies of the locally $D$-optimal designs for the different models.}
	\label{tabdeff}
\end{table*}

	 Note that the selected models for the four responses differ and it is not clear if a locally
  $D$-optimal design for a particular model (for example
  the model used for temperature) has good properties in the models used for the other responses. In Table \ref{tabdeff} we show the $D$-efficiencies if a locally $D$-optimal design for a particular model is used for a different model. We observe a substantial loss of efficiency. For example if the locally $D$-optimal design for the flame intensity is used its $D$-efficiency for analyzing the flame width is only $65.77\%$.
In order to address this problem  we  have used the
  multi-objective criterion \eqref{Bday} to find a design $\mathbf{X}^{(2)}$ for the observations on a different day with reasonable $D$-efficiencies in all models under consideration.  We begin considering only uncertainty with respect to the model in the criterion
  \eqref{eff}, while all the parameters are fixed.
  We used equal weights
  for all four  models from Table \ref{tab:Properties} as prior distribution and the resulting design is given in the left part of Table \ref{multiobj}.\\
The corresponding efficiencies
  \begin{equation}\label{refeff}
   \mbox{eff}_D (\mathbf{X}^{(2)}_R,\mathbf{X}_B^{*(2)}|s) =  \Bigl( {| I(\mathbf{X}_R,s)|  \over |I(\mathbf{X}^*_B,s)|} \Bigr)^{1/(p(s)+1)}
\end{equation}

 \begin{equation}\label{refeffD1}
   \mbox{eff}_{D_1} (\mathbf{X}^{(2)}_R,\mathbf{X}_B^{*(2)}|s) =  \frac {\Phi_{D_1}(\mathbf{X}_R^{(2)}|s)}{\Phi_{D_1}(\mathbf{X}^{*(2)}_B|s)}
\end{equation}
of the reference design $\mathbf{X}_R=(\mathbf{X}^{(1)}_C, \mathbf{X}_R^{(2)})$  with respect to the Bayesian $D$-optimal design $\mathbf{X}^*_B= (\mathbf{X}_C^{(1)}, \mathbf{X}^{*(2)}_B ) $ are presented
  in the second line of Table  \ref{efficlocDopt} and \ref{Ddayeff}, respectively. We observe a similar improvement with respect to $D$-efficiency as obtained by the locally $D$-optimal designs.
  From this table we can also easily calculate the $D$-efficiencies of the design ${\mathbf{X}^*_B}^{(2)}$, which are given by $98.04\%$, $97.52\%$,  $91.45\%$, $92.77\%$
  in the models for the temperature, velocity, flame width and flame intensity, respectively. Similarly, the efficiencies $\mbox{eff}_{D_1}(\mathbf{X}^{*(2)}_B, \mathbf{X}^{*(2)}_s)$ of the design ${\mathbf{X}^*_B}$ with respect to the locally $D$-optimal designs for estimating the parameter $\gamma$ are obtained as $132.53\%$, $66.05\%$, $111.90\%$, $254.67\%$. This means that the compromise improves the reference design in all models with respect to parameter estimation. On the other hand for the estimation of the day effect only an improvement in the model for flame width is achieved (note that the design ${\mathbf{X}^*_B}^{(2)}$ is not constructed for this purpose). 
\\
  While rather precise information is available for the parameter $\beta$ from the first $30$ observations,
  the designs and its properties might be sensitive with respect to the specification of the parameter $\gamma$ for the additional
  day effect. In order to construct designs, which address this uncertainty we can also use the criterion \eqref{eff}, where we now
  also allow for uncertainty with respect to the parameter $\gamma$ in the criterion. More precisely, for each of the four models
  we consider three possible values for $\gamma$, namely the value used in the local $D$-optimality criterion and $90\%$
  and $110\%$ of this value (for example for the temperature model we used $14.4$, $16$, and $17.6$ as possible values of $\gamma$). 
      \begin{table}[!t]
\begin{center}
{\scriptsize
\begin{tabular}{ccccc|| cccc}
  \hline
 Run & $L$ & $K$ & $D$ & $FDV$  & $L$ & $K$ & $D$ & $FDV$\\
  \hline
1 & 2& 2 &2& 2 			 &2 & 2 & 2  & -2 \\
  2 & 2 & -2 &2& -2 			&2 & -2 & -2 &-2  \\
  3 & -2 &0.34 & -2 & 2 		 &-2 & 0.37 & -2 & 2\\
  4 & -2 &-2 &-2& -2			  &-2 &-2 &2 & 2\\
   \hline
\end{tabular}}
\end{center}
\vspace{-.5cm}
\caption{ \it Bayesian  $D$-optimal designs with respect to the criterion \eqref{Bday} for the four generalized linear models specified in Table \ref{tab:Properties}. Left part: parameter of the day effect $\gamma$ is fixed; right part:
three values for the parameter of the day effect $\gamma$,  $\gamma \pm 10\%\gamma$.}
\label{multiobj}
\end{table}

  The resulting criterion
  \eqref{eff} therefore consists of a sum of $12$ terms and the maximizing design is depicted in the right part of Table \ref{multiobj}. The structure of the two Bayesian $D$-optimal designs is very similar, since both designs put most of the design points in the edges of the design space. The $D$- and $D_1$-efficiencies are presented in the third rows of Table \ref{efficlocDopt} and \ref{Ddayeff}, respectively. Because of the similarity of the two Bayesian $D$- optimal designs the efficiences have nearly the same values.\\
These investigations show that the $D$-optimal designs yield a substantial improvement of the reference design if all parameters in the model \eqref{daymod} have to be estimated. On the other hand, if the only interest of the experiment is the estimation of a day effect, the reference design yields a more precise estimate of the parameter $\gamma$ in the models for temperature, velocity and flame intensity than optimal designs based on $D$-optimality criteria.

  \subsection{Optimal designs for testing for an additional day effect} \label{sec52}

 If the main interest of the experiment is the investigation of the existence of an additional day effect the design can be constructed such that the test for the hypothesis
 $H_0:\gamma=0$  is most powerful, which is reflected by the criterion $\Phi_{D_1}$ defined in  \eqref{dopt1}. The corresponding multi-objective criterion addressing
 uncertainty with respect to the regression model, link function and parameters is given by \eqref{critbayD1}. The locally $D_1$-optimal designs for the four models
 in Table \ref{tab:Properties} are presented in Table \ref{tab:locoptD1} in  Appendix \ref{secdesigns2}. We observe that in contrast to $D$-optimal designs $D_1$-optimal designs do not use the edges of the design space. The efficiencies of the reference designs $\mathbf{X}^{(2)}_R$
 are given in the first row of Table \ref{efficlocD1opt}. For the temperature and velocity the $D_1$-efficiencies of the reference design are about $90\%$. On the other hand an improvement of the reference designs can be observed for velocity and flame width (here the efficiencies are $63.81\%$ and $86.51\%$, respectively).

  \begin{table*}
	\centering
	{\scriptsize
		\begin{tabular}{|c|c|c|c|}
		\hline
			  temperature & velocity & flame width & flame intensity \\
			 \hline
			 \hline
			90.51\% & 86.51\% & 63.81\% & 90.85\% \\
			90.55\% & 86.61\% & 64.68\%  &94.70\% \\
			90.64\% & 86.69\% & 66.05\% & 94.37\% \\
			\hline
%			Link	& identity & logistic & inverse & identity\\
%			\hline
		\end{tabular}}
		\vspace{-.1cm}
	\caption{\it First row:  $D_1$-efficiencies of the reference design. Second row: $D_1$-efficiencies of the reference design $\mathbf{X}^{(2)}_R$
	with respect  to the design $\mathbf{X}_{B_1}^{*(2)}$ maximizing the multi objective criterion \eqref{critbayD1}, where $\gamma$ has been fixed.
	Third row: $D_1$-efficiencies of the reference design $\mathbf{X}^{(2)}_R$ with respect to
	the design $\mathbf{X}_{B_1}^{*(2)}$ maximizing the multi objective criterion \eqref{critbayD1}, where uncertainty with respect to the parameter $\gamma $ has been addressed.}
	\label{efficlocD1opt}
\end{table*}
 As in the previous section we construct a robust  design for testing for an additional day effect by
  maximizing the multi objective criterion \eqref{critbayD1}, where all parameters have been fixed
 ($\beta$ is obtained from Table \ref{tab:bmtemai},
 while information from other experiments was used for  the parameter $\gamma$ for the construction of the locally optimal designs, that is  $\gamma =-16$, $\gamma =0.01$, $\gamma= 0.002$ and $\gamma= 0.09$
 in the models  for  temperature, velocity, flame width and flame intensity, respectively). The resulting design is shown in the left part of Table \ref{multiobjD1} and its
 efficiencies are presented in the second row of Table \ref{efficlocD1opt}. We observe a similar improvement of the reference designs as obtained by the locally $D_1$-optimal designs. Finally, we consider designs addressing the fact that the parameter $\gamma $ cannot be estimated from the data of the initial day. If we address the
 uncertainty about this parameter in the same way as described in the previous section we obtain the design presented in the right part of Table \ref{multiobjD1}. The $D_1$-efficiencies of the
 reference designs $\mathbf{X}^{(2)}_R $ with respect to this design are shown in the third row of Table  \ref{efficlocD1opt}.\\
Both Bayesian $D_1$-optimal designs are similar but differ substantially from the two Bayesian $D$-optimal designs in Table \ref{multiobj}. The $D_1$-optimal designs use more experimental conditions from the interior of the design space $[-2,2]^4$. Their efficiencies are very similar with respect to the locally $D_1$-optimal designs and range between $65\%$ and $95\%$. Whereas the reference design performs nearly as well as the two Bayesian $D_1$-optimal designs in the cases of temperature and velocity, in the cases of flame width and flame intensity the Bayesian $D_1$-optimal yields more precise estimates as to the reference designs. \\
On the other hand the $D$-efficiencies of the  Bayesian $D_1$-optimal designs are given by $85.14\%$,  $73.69\%$, $79.58\%$ and $69.23\%$ for the temperature
velocity, flame width and flame intensity, respectively, and therefore these designs are not  very efficient for estimating
all parameters in a generalized linear model with an additional day effect.
  \begin{table}[!t]
\begin{center}
{\scriptsize
\begin{tabular}{ccccc|| cccc}
  \hline
 Run & $L$ & $K$ & $D$ & $FDV$  & $L$ & $K$ & $D$ & $FDV$\\
  \hline
1 &  0.03 & -1.62 & 2.00 & -0.65 		& -0.57 & 0.13 & -1.15 & -0.96  \\
  2 & 0.90 & 0.36 & 0.44 & -0.57	 & 0.46 & -1.53 & 1.97 & -0.61 \\
  3 & 1.11 & 0.53 & -2.00 & -0.70 	&-1.37 & 0.45 & -0.61 & 1.83 \\
  4 & -1.83  & 0.54 &-0.53 & 2.00 &  1.42&  0.84 & -0.27 & -0.60 \\

   \hline
\end{tabular}}
\end{center}
\vspace{-.5cm}
\caption{ \it Bayesian  $D_1$-optimal designs with respect to the criterion \eqref{critbayD1} for the four generalized linear models specified in Table \ref{tab:Properties}. Left part: parameter of the day effect $\gamma$ is fixed; right part:
three values for the parameter of the day effect, $\gamma$,  $\gamma \pm 10\%\gamma$.}
\label{multiobjD1}
\end{table}

  \subsection{Efficient designs for estimating  and testing } \label{sec53}

The numerical results of Section \ref{sec51} and   \ref{sec52} show that different objectives
such as estimation of all parameters and testing for an additional day effect result in rather different
experimental designs, and  optimal  designs for one particular task (such as maximization of the power) are usually not
efficient for the other (estimation of parameters). In order to construct efficient designs for these contradicting
tasks we consider in a final step a compromise criterion of the form
 \begin{equation} \label{comp}
\alpha \Phi_{B} (\mathbf{X^{(2)}}|\mathcal{M}) + (1- \alpha ) \Phi_{B_1} (\mathbf{X^{(2)}}|\mathcal{M}),
\end{equation}
where $\Phi_{B} $ and $ \Phi_{B_1}$ are defined in \eqref{Bday} and \eqref{critbayD1}, respectively, and $\alpha \in [0,1]$ is a
pre-determined constant reflecting
the importance of the different goals estimation and testing. The resulting design for four additional experiments is shown in Table \ref{tab:descombicrit}
for $\alpha =0.5$, where we use in both criteria the same prior distributions as described in Section \ref{sec51} and   \ref{sec52}.
The  corresponding efficiencies are depicted in Table \ref{tabkombieff} and we  observe that this design yields  high $D_1$-efficiencies
in all four models under consideration and additionally  a substantial improvement with respect to  the $D$-efficiencies, which  vary between $76\%$
and $99\%$.  The $D$-efficiencies could be increased if larger values of $\alpha $ are used in the criterion \eqref{comp} at the expense
 of smaller $D_1$-efficiencies. For the sake of brevity these results are not depicted.
\begin{table}[t]
\begin{center}
{\scriptsize
\begin{tabular}{ccccc}
  \hline
 Run & $L$ & $K$ & $D$ & $FDV$ \\
  \hline
1 &0.10  &0.17  &  2.00  & -0.76 \\
  2 &0.29 & -2.00 &  -2.00 & -1.05\\
  3 & 1.75 &  0.58  &  2.00 & -0.08 \\
  4 &  -2.00   & 0.41 &   -2.00 &  2.00 \\

   \hline
\end{tabular}}
\end{center}
\vspace{-.5cm}
\caption{\it The compromise design maximizing the criterion \eqref{comp}.The Bayesian criteria  $\Phi_{B} $ and $ \Phi_{B_1}$ only consider model uncertainty.}
\label{tab:descombicrit}
\end{table}

\begin{table}[t]
	\centering
	{\scriptsize
		\begin{tabular}{|c||c|c|c|c|}
		\hline
		 & \multicolumn{4}{c|}{model}\\
		 \cline{2-5}
			  & temperature & velocity & flame width & flame intensity \\
			 \hline
			 \hline
			$D$-efficiency & 91.15\% &77.84\%& 85.38\%&76.12\% \\
			\hline
			$D_1$-efficiency & 99.17\% & 98.50\% & 97.89\% & 91.02\%\\
			\hline
		\end{tabular}}
		\caption{\it $D$- and $D_1$-efficiences of the compromise design maximizing the criterion \eqref{comp}. }
	\label{tabkombieff}
\end{table}

\subsection{Experimental results}
A further series of  eight experiments  was conducted to  improve the understanding of  the thermal spraying process, where
four runs were performed  under a reference and an optimal design, respectively.
Because the goal  was to estimate all parameters in the models for temperature, velocity, flame width and intensity
(including the day effect) a Bayesian $D$-optimal design was calculated.
In order to address the uncertainty with respect to the day effect $\gamma$
 an average over five possible values for $\gamma$ was calculated, i.e. $\gamma$,  $\gamma \pm 10\%\gamma$, $\gamma \pm 20\%\gamma$. 
 The calculated reference and Bayesian $D$-optimal design are depicted in Table \ref{opt4}. The 
Bayesian $D$-optimal design  advices the experimenter to use experimental conditions at the boundary of the design space 
whereas the reference design naturally stays more towards the center (i.e. it only contains coded $-1$, $1$ of the input parameters). \\
 The observed data are given in Table \ref{locd1four} and \ref{locd1ref} in Appendix \ref{secdesigns3} and it can be seen that 
 the particle properties at the extreme experimental conditions differ substantially from the  original data. 
 For example, in run number $2$ the  temperature  and the value of flame intensity  are  only  $1298.81$  and $9.84$, respectively.
 As a consequence adding the four additional runs to the initial data leads to noticeable changes in the parameter estimates for the model 
 of temperature if the optimal design is used. All other parameter estimates are only marginally altered
 (these results are not displayed for the sake of brevity). \\
 The $D$-efficiencies \eqref{refeff}
 of the reference design with respect to this Bayesian $D$-optimal design are
 displayed in the first row of Table \ref{tab:mse} and are 
  given by $79.53\%$, $75.35\%$,  $78.14\%$ and $69.03\%$.  It is also 
  of interest to compare these ''theoretical'' values 
  (based on the $30$ initial  observations) with the ''observed'' $D$-efficiencies 
    (based on  the estimated covariance matrices from the   $30$ initial  plus  four additional observations), which are
    shown in the second line of Table \ref{tab:mse}.  These results show a substantial  improvement between 
    $39\%$ (temperature) and $86\%$ (flame width) with respect to the $D$-criterion, if the $4$ additionals
    runs are performed according to the Bayesian $D$-optimal design.  \\
\begin{table}[htpb]
\begin{center}
{\scriptsize
\begin{tabular}{ccccc|cccc}
  \hline
 Run & $L$ & $K$ & $D$ & $FDV$  & $L$ & $K$ & $D$ & $FDV$   \\
  \hline
1  &  2  &  2       &  2  &  -0.53 & 1&  1& -1&  -1  \\
2  &  -2 &  -2      &  2  &     -2   & -1&  1&  1&  -1 \\
3  &  -2 &  0.31 &  -2 &      2   & 1&  1&  1&   1  \\
4  &  2  &  -2      &  -2 &     -2    &1& -1&  1&  -1  \\
   \hline
\end{tabular}}
\end{center}
%\vspace{-0.5cm}
\caption{\it Right part: Bayesian  $D$-optimal design with respect to the criterion \eqref{Bday} with five values for the parameter of the day effect $\gamma$,  $\gamma \pm 10\% \gamma$, $\gamma \pm 20\%$. Left part: The reference design. }
\label{opt4}
\end{table}
\begin{table}[htpb]
\begin{center}
{\scriptsize
\begin{tabular}{|c||c|c|c|c|}
  \hline
              & Temperature & Velocity & Flame Width & Flame Intensity \\
  \hline
theoretical $D$-efficiency   & 79.53\%   & 75.35\% &  78.14\%     & 69.34\%            \\
observed $D$-efficiency   & 39.34\%    & 85.89\% &  86.46\%     & 57.05\%             \\
\hline\hline
MSE   optimal  design & 27.40     & 12.65  &  1.43     & 1.89        \\
MSE reference design   & 16.97    & 8.91  &  4.10     & 2.73          \\
  \hline
\end{tabular}}
\end{center}
%\vspace{-0.5cm}
\caption{ \it Upper part: $D$-efficiencies (theoretical and observed)  of the reference design with respect to the Bayesian $D$-optimal design. 
Lower part:
MSE for the prediction of  $14$  runs from (\ref{tab:addruns})  using four additional observations 
from an optimal and  a reference design.}
\label{tab:mse}
\end{table}
It might also be of interest to compare both designs with respect to their prediction properties.   On the same day of the eight  new experiments, 
fourteen additional  experiments were conducted with different aims. The design and data are shown on Table \ref{tab:addruns}. We 
 investigate  how the estimated  generalized linear models from the two designs perform 
with respect to  prediction of these  measured in flight properties. 
The  lower part of  Table \ref{tab:mse}  presents  the mean squared  error between predicted and measured particle properties, and 
we observe that the reference design leads to smaller mean squared errors for temperature and velocity 
whereas the Bayesian $D$-optimal design yields better predictions of the  flame width and intensity.

\section{Concluding remarks}
\def\theequation{5.\arabic{equation}}
\setcounter{equation}{0}

In this paper we have investigated optimal designs for analyzing thermal spraying processes on the basis of generalized linear models,
where observations are available from experiments conducted at two different days. While a central  composite
design is used for  the experiments on the first day, optimal designs for the experiments on a further day are  constructed,
which also allow for testing the existence of an additional additive day effect in the generalized linear model.
Uncertainty with respect to the model assumptions occurs from several perspectives and is addressed in the optimality criteria
used for the construction of efficient designs. Firstly, one design is constructed for analyzing various  in-flight properties as
temperature,    velocity,  flame width   and   flame intensity with different generalized linear models. Secondly, the criteria
also address the problem uncertainty with respect to unknown day effect.

We consider  $D$- and $D_1$-optimality criteria which yield designs  minimizing the volume of the ellipsoid of concentration
for the vector of all parameters and maximizing the power of the likelihood ratio test for the existence of an additional
day effect, respectively.
Bayesian $D$- and $D_1$-optimal designs  (addressing the problem of model uncertainty and  imprecise information about the unknown day effect)
are determined and investigated with respect to their statistical efficiencies. $D$-optimal designs use the edges of
the design space and it is demonstrated that these designs improve a reference design substantially with respect to the efficiency for
estimating all parameters in the generalized linear model. On the other hand, in many cases the reference design yields more power for the
likelihood ratio test of an additional day effect than the $D$-optimal designs and the reference design can be improved by a Bayesian
$D_1$-optimal design. Therefore,
if the only goal of the additional experiments is the investigation of an additional day effect Bayesian $D_1$-optimal designs should
be used. These designs advice the experimenter to take more observations in the interior
of the design space. Thus the two objectives estimation of all parameters and testing for an additional day effect result in rather different
experimental designs and the goals of the experiment have to be carefully defined before optimal designs are constructed for
the analysis of thermal spraying processes with generalized linear models. If this is not possible, a compromise design criterion can
be developed, which yields designs with reasonable efficiencies for estimating all parameters  and testing for an
additional day effect  under model uncertainty. \\
Finally, we use the results of this paper to design new experiments for the analysis of the thermal spraying process
and demonstrate that a Bayesian $D$-optimal design improves a reference design with respect to the $D$-optimality criterion.
On the other hand, for the prediction of $14$ additional experiments the superiority of the Bayesian $D$-optimal design 
is only visible for the responses flame width  and  intensity.   
%
%However, as no final conclusions can be drawn from just one practical run of the design and  optimal designs showed better theoretical results  their use will be 
% further followed up in the thermal spraying application.
\medskip
\medskip
\medskip

{\bf Acknowledgements} The authors would like to thank Martina
Stein, who typed parts of this manuscript with considerable
technical expertise.
This work has been supported in part by the
Collaborative Research Center "Statistical modeling of nonlinear
dynamic processes" (SFB 823, Projects B1 and C2) of the German Research Foundation
(DFG).

{  \small
\bibliographystyle{apalike}
 \setlength{\bibsep}{ 2pt}
\bibliography{gamma}
}

\newpage
			%%%%%%%%%%%%%%%%%%%%%%%%%%%%%%%%%%%%%%%%%%%%%%%%%%%%%%%%%%%%%%%%%%%%%%%
			%%%%%%%%%%%%%%%%%%%%%%%%%%%%%%%%%%%%%%%%%%%%%%%%%%%%%%%%%%%%%%%%%%%%%%%
			%%%%%%%%%%%%%%%%%%%%%%%%%%    Appendix	   %%%%%%%%%%%%%%%%%%%%%%%%%%%%
			%%%%%%%%%%%%%%%%%%%%%%%%%%%%%%%%%%%%%%%%%%%%%%%%%%%%%%%%%%%%%%%%%%%%%%%

\appendix
\section{Appendix: Data, designs and  and estimates}

\subsection{Estimates in the identified models and the standard design}  \label{secbestfits}

In this section we display the parameter estimates in the models identified by the BIC
for the four responses. The values are obtained from the $30$ observations of the  initial day
and are used in the local optimality criteria to construct the optimal design for the additional
four runs on the next day.

\def\theequation{A.\arabic{equation}}
\setcounter{equation}{0}
\nopagebreak
\begin{table}[htpb]
\begin{center}
{\scriptsize
\begin{tabular}{ccccc||ccccc||ccccc}
\hline
Run &  L  &  K  &  D  & FDV & Temperature & Velocity & Flame Width & Flame Intensity  \\
\hline
1   &  1 & -1  &  1   &  -1   &  1450.5706   & 674.1324  &  7.9059 & 13.1971  \\
2   &  1 &  1  &  1   &   1   &  1500.9382   & 726.6706  & 12.4912 & 21.0029  \\
3   & -1 & -1  &  1   &  -1   &  1484.8952   & 649.1190  &  8.1238 & 15.3929  \\
4   & -1 & -1  & -1   &   1   &  1534.6750   & 666.0781  & 13.5563 & 21.4375  \\
5   &  0 &  0  &  0   &   0   &  1519.4829   & 709.3029  & 11.9629 & 19.7143  \\
6   &  0 &  0  &  0   &   0   &  1527.6065   & 713.6581  & 12.1742 & 19.9419  \\
7   & -1 &  1  &  1   &  -1   &  1543.3053   & 730.3474  & 10.3711 & 18.3579  \\
8   & -1 &  1  & -1   &   1   &  1574.0970   & 739.4212  & 14.9909 & 23.3667  \\
9   &  1 &  1  & -1   &   1   &  1536.2371   & 756.7057  & 13.7657 & 21.8543  \\
10  &  1 & -1  & -1   &  -1   &  1497.6209   & 698.4093  &  8.7767 & 15.8093  \\
11  &  0 &  0  &  0   &   0   &  1527.8571   & 710.8250  & 11.9821 & 19.9393  \\
12  & -1 &  1  & -1   &  -1   &  1564.3114   & 753.5943  & 11.1229 & 18.7143  \\
13  &  1 &  1  & -1   &  -1   &  1528.9267   & 770.7367  &  9.5000 & 17.0000  \\
14  & -1 &  1  &  1   &   1   &  1546.6594   & 714.0031  & 14.8187 & 23.5625  \\
15  &  1 & -1  &  1   &   1   &  1484.7806   & 665.0000  & 12.3472 & 20.5139  \\
16  & -1 & -1  &  1   &   1   &  1502.0265   & 640.9088  & 13.2176 & 21.3500  \\
17  & -1 & -1  & -1   &  -1   &  1525.3917   & 678.9194  & 10.0417 & 17.2917  \\
18  &  1 &  1  &  1   &  -1   &  1508.2706   & 749.0647  &  8.5206 & 15.9706  \\
19  &  0 &  0  &  0   &   0   &  1535.5706   & 714.2500  & 12.3294 & 20.1412  \\
20  &  1 & -1  & -1   &   1   &  1504.6000   & 689.5364  & 12.6121 & 20.2879  \\
21  &  0 &  0  &  0   &   0   &  1521.7227   & 708.9636  & 11.7977 & 19.6568  \\
22  &  0 &  0  & -2   &   0   &  1534.7182   & 726.6697  & 11.7939 & 19.2697  \\
23  & -2 &  0  &  0   &   0   &  1542.8600   & 688.1171  & 12.4971 & 20.2914  \\
24  &  2 &  0  &  0   &   0   &  1462.0088   & 723.5471  &  9.1765 & 16.7735  \\
25  &  0 &  0  &  0   &   0   &  1521.4765   & 709.1412  & 11.5176 & 19.2412  \\
26  &  0 &  0  &  0   &  -2   &  1516.5378   & 708.6919  & 11.3649 & 19.0757  \\
27  &  0 &  0  &  2   &   0   &  1491.7684   & 684.3026  & 10.4868 & 18.5632  \\
28  &  0 &  2  &  0   &   0   &  1512.7982   & 755.1382  & 10.8436 & 18.8527  \\
29  &  0 &  0  &  0   &   2   &  1520.6485   & 695.1848  & 14.4455 & 22.7879  \\
30  &  0 & -2  &  0   &   0   &  1435.7488   & 612.6093  &  8.9209 & 16.0163  \\
\hline
\end{tabular}}
\end{center}
\vspace{-0.5cm}
	\caption{\it Central Composite Design used for the first $30$ observations}
	\label{tab:CCD}
\end{table}

\begin{table}[htpb]
\begin{center}
{\scriptsize
\begin{tabular}{ccccc}
  \hline
               & Temperature & Velocity & Flame Width & Flame Intensity \\
  \hline
(Intercept)    & 1523.2627 (2.6722) &  6.5648  (0.0016) &  0.0863  (0.0018)  &  19.4784 (0.3364) \\
  $L$          &  -17.7423 (2.3136) &  0.0136  (0.0014) &  0.0053  (0.0015)  &  -0.8887 (0.1901) \\
  $K$          &   19.6580 (2.2939) &  0.0516  (0.0014) & -0.0044  (0.0016)  &   0.8646 (0.1863) \\
  $D$          &  -13.8181 (2.3136) & -0.0171  (0.0014) &  0.0029  (0.0015)  &  -0.3709 (0.1970) \\
  $FDV$        &   -                & -0.0078  (0.0014) & -0.0123  (0.0015)  &   2.1661 (0.2042) \\
  $L^2$        &   -                &   -               &   -                &  -0.3096 (0.1760  \\
  $K^2$        &   -9.9897 (2.0813) & -0.0092  (0.0012) &  0.0039  (0.0015)  &  -0.5615 (0.1925) \\
  $FDV^2$      &   -                &   -               &   -                &   0.5092 (0.1699) \\
  $L\cdot K$   &   -                & -0.0031  (0.0017) &   -                &    -              \\
  $D\cdot FDV$ &   -                &   -               &   -                &   0.4095 (0.2378) \\
   \hline
\end{tabular}}
\end{center}
\vspace{-0.5cm}
\caption{\it Parameter estimates and standard errors in brackets of the models for temperature, velocity, flame width and flame intensity chosen by the BIC.}
\label{tab:bmtemai}
\end{table}

\newpage
\subsection{Optimal designs with four runs} \label{secdesigns2}

\begin{table}[htpb]
\begin{center}
{\scriptsize
\begin{tabular}{cccc||cccc||cccc||cccc}
  \hline
   & \multicolumn{3}{c||}{temperature}& \multicolumn{4}{c||}{velocity} & \multicolumn{4}{c||}{flame width}& \multicolumn{4}{c}{flame intensity}\\
  \hline
 Run & $L$ & $K$ & $D$ &$L$ & $K$ & $D$ & $FDV$ & $L$ & $K$ & $D$ & $FDV$ &$L$ & $K$ & $D$ & $FDV$ \\
  \hline
  1 & -2 &-0.07 & -2 &      -2 & 2 &-2 & 2 			&2 & 0.37 & 2 & 2 & 			2 &2 & 2 &-2 \\
  2 &  2 & -0.13 & 2 &   - 2 & -2 &2 & 2  & -2 & 2 & -2 &  2 &				 -2 &-2 & 2 &-2\\
  3 &2 & 2& -2  &                     2 & 2 & -2 &  -2 & -2 & 0.08 & -2 & 2 & 				0.47&-0.95 & 2 & -0.64  \\
  4 & -2 & -2 & 2 &    2 & -2 & 2 & -2 & 2 & 0.37 & -2 & -2			 & 2&-2 & -2&  -2  \\
\hline
\end{tabular}}
\end{center}
\vspace{-0.5cm}
\caption{ \it Locally D-optimal designs for the responses temperature (left part), velocity (middle left part), flame width (middle right part) and flame intensity (right part) \label{tab:locoptD} }
\end{table}

\begin{table}[htpb]
\begin{center}
{\scriptsize
\begin{tabular}{cccc|| cccc || cccc|| cccc}
    \hline
   & \multicolumn{3}{c||}{temperature}& \multicolumn{4}{c||}{velocity} &\multicolumn{4}{c||}{flame width} & \multicolumn{4}{c}{flame intensity}  \\
  \hline
 Run & $L$ & $K$ & $D$ &		  $L$ & $K$ & $D$ & $FDV$& 	$L$ & $K$ & $D$ & $FDV$  & 				$L$ & $K$ & $D$ & $FDV$\\
  \hline
1 & -1.23 &-1.32 & -0.05&		0.00 & 0.96 & 0.64 & -0.73 	&  -1.02 & 0.29 & 1.72 & -1.83 & 			1.59 & 1.03 & 1.30 & -1.075 \\
2 & 0.10 &  0.94 &  0.81 &			 -0.54 & -0.62 & -1.90 		& 0.84 &-1.81 &  0.99 & 0.29 & 1.92 &		   0.09 & -1.53 & 1.43 & -0.51\\
3 & 0.28 &  0.64 & 0.42 &			  0.40 &  -1.12 & 1.31 		& -1.25  &   0.07 & -1.07 & -1.40  & -0.29 &	0.00 & -0.61 & -0.66 & 0.64\\
4 & 1.20 & -0.64 & -0.89 &		 0.13 & 0.79 & -0.05  &1.14 	& 1.54 &0.28 &-1.58 &1.50&		   		-0.85 - & 0.22 & -1.79  & -1.11\\
   \hline
\end{tabular}}
\end{center}
\vspace{-0.5cm}
\caption{ \it Locally $D_1$-optimal designs for the responses temperature (left part), velocity (middle left part),  flame width (middle right part) and flame intensity (right part) \label{tab:locoptD1} }
\end{table}

\newpage
\subsection{Bayesian $D$-optimal and reference design for $4$ additional runs} \label{secdesigns3}

\begin{table}[htpb]
\begin{center}
{\scriptsize
\begin{tabular}{ccccc|| cccc}
  \hline
 Run & $L$ & $K$ & $D$ & $FDV$  & Temperature  & Velocity  & Flame Width & Flame Intensity \\
  \hline
1  &  2  &  2       &  2  &  -0.53 & 1466.8123 &  787.5585  &  12.3446   &  22.0938 \\
2  &  -2 &  -2      &  2  &     -2    & 1298.8123 &  593.2692  &  10.2261   &   9.8646 \\
3  &  -2 &  0.31 &  -2 &      2    & 1560.3545 &  706.1242  &  18.9030   &  28.6333 \\
4  &  2  &  -2      &  -2 &     -2    & 1437.7284 &  687.5351  &   7.4500   &  13.8446 \\
   \hline
\end{tabular}}
\end{center}
%\vspace{-0.5cm}
\caption{\it Bayesian $D$-optimal design for four additional runs}
\label{locd1four}
\end{table}
\begin{table}[h!]
\begin{center}
{\scriptsize
\begin{tabular}{rrrrr|| rrrr}
  \hline
 Run & $L$ & $K$ & $D$ & $FDV$  & Temperature  & Velocity  & Flame Width & Flame Intensity \\
  \hline
1 & 1&  1& -1&  -1&    1527.1426 & 778.2632 & 13.2456 & 22.3985  \\
2 &-1&  1&  1&  -1&    1493.9143 & 752.6063 & 12.4841 & 22.3333  \\
3 & 1&  1&  1&   1&    1507.5667 & 752.8273 & 17.6909 & 27.9348  \\
4 & 1& -1&  1&  -1&    1443.8103 & 696.7851 & 10.4471 & 18.8977  \\
   \hline
\end{tabular}}
\end{center}
%\vspace{-0.5cm}
\caption{\it Reference design  for four additional runs}
\label{locd1ref}
\end{table}

\subsection{Additional runs and results} \label{results}

\begin{table}[htpb]
\begin{center}
{\scriptsize
\begin{tabular}{rrrrr|| rrrr}
  \hline
 Run & $L$ & $K$ & $D$ & $FDV$  & Temperature  & Velocity  & Flame Width & Flame Intensity \\
  \hline
1 & 0.01 &  1.09 & -0.20 & -1.67   &  1514.6610 & 782.1146 &  10.2951 & 19.1976 \\
2 & 0.01 &  1.09 & -0.20 & -1.67   &  1521.8186 & 786.4209 &  10.0116 & 18.6930 \\
3 & 1.82 & -0.36 &  0.46 & -0.58   &  1475.2022 & 734.7978 &  11.7778 & 21.1467 \\
4 & 1.82 & -0.36 &  0.46 & -0.58   &  1488.6825 & 737.3925 &  11.1850 & 20.0250 \\
5 & 1.27 & -1.32 &  0.38 & -0.71   &  1434.2327 & 687.7714 &  10.4510 & 18.9408 \\
6 & 1.27 & -1.32 &  0.38 & -0.71   &  1456.8717 & 689.5453 &  10.0019 & 17.7623 \\
7 & 0.00 & -0.01 &  0.20 & -1.73   &  1478.7136 & 743.1182 &   9.7227 & 17.1773 \\
8 & 0.00 & -0.01 &  0.20 & -1.73   &  1520.9761 & 747.9326 &   9.3457 & 16.3413 \\
9 &-0.48 &  0.50 & -0.60 &  1.78   &  1529.3061 & 726.5163 &  17.9367 & 27.6449 \\
10&-0.48 &  0.50 & -0.60 &  1.78   &  1521.4094 & 721.4434 &  17.4113 & 27.1906 \\
11& 1.00 & -1.00 & -1.00 &  1.00   &  1507.4579 & 698.7368 &  16.2667 & 25.5053 \\
12&-1.00 & -1.00 & -1.00 & -1.00   &  1491.3108 & 696.6215 &  12.4585 & 21.0092 \\
13&-1.00 & -1.00 &  1.00 &  1.00   &  1453.4762 & 661.9381 &  16.2333 & 26.2127 \\
14&-1.00 &  1.00 & -1.00 &  1.00   &  1552.7875 & 749.2734 &  18.3484 & 27.9703 \\
   \hline
\end{tabular}}
\end{center}
%\vspace{-0.5cm}
\caption{\it Additional runs used for the investigation of the prediction properties of the reference and
optimal design}
\label{tab:addruns}
\end{table}

\end{document}